\documentclass[reprint,aip]{revtex4-2}
\usepackage[utf8]{inputenc}
\usepackage{xprintlen}
\usepackage{amsmath}
\usepackage{physics}
\usepackage{mhchem}
\usepackage{url}
\usepackage{amsfonts}
\usepackage{amssymb}
\usepackage{tabularx}
\usepackage{mathtools}
\usepackage{subfigure}
\usepackage{booktabs}
\usepackage{makeidx}
\usepackage{graphicx}
\usepackage{lmodern}
\usepackage{siunitx}
\usepackage{upgreek}
\usepackage{adjustbox}
\usepackage{blindtext}
\usepackage{siunitx}
\usepackage[section]{placeins}
\usepackage{float}
\usepackage{braket}
\usepackage{appendix}
\usepackage{xr}
\begin{document}

\title{Optimal control of a nitrogen-vacancy spin ensemble in diamond for sensing in the pulsed domain}
\author{Andreas F.L. Poulsen}
\thanks{These authors contributed equally to this work}
\affiliation{Center for Macroscopic Quantum States (bigQ), Department of Physics, Technical University of Denmark, Kgs. Lyngby, Denmark}%
\author{Joshua D. Clement}%
\thanks{These authors contributed equally to this work}
\affiliation{Center for Macroscopic Quantum States (bigQ), Department of Physics, Technical University of Denmark, Kgs. Lyngby, Denmark}%
\author{James L. Webb}
\email{jaluwe@fysik.dtu.dk}
\affiliation{Center for Macroscopic Quantum States (bigQ), Department of Physics, Technical University of Denmark, Kgs. Lyngby, Denmark}%
\author{Rasmus H. Jensen}
\affiliation{Center for Macroscopic Quantum States (bigQ), Department of Physics, Technical University of Denmark, Kgs. Lyngby, Denmark}%
\author{Kirstine Berg-S{\o}rensen}
\affiliation{Department of Health Technology, Technical University of Denmark, Kgs. Lyngby, Denmark}%
\author{Alexander Huck}
\email{alexander.huck@fysik.dtu.dk}
\affiliation{Center for Macroscopic Quantum States (bigQ), Department of Physics, Technical University of Denmark, Kgs. Lyngby, Denmark}%
\author{Ulrik Lund Andersen}
\email{ulrik.andersen@fysik.dtu.dk}
\affiliation{Center for Macroscopic Quantum States (bigQ), Department of Physics, Technical University of Denmark, Kgs. Lyngby, Denmark}%

\begin{abstract}
Defects in solid state materials provide an ideal, robust platform for quantum sensing. To deliver maximum sensitivity, a large ensemble of non-interacting defects hosting coherent quantum states are required. Control of such an ensemble is challenging due to the spatial variation in both the defect energy levels and in any control field across a macroscopic sample. In this work we experimentally demonstrate that we can overcome these challenges using Floquet theory and optimal control optimization methods to efficiently and coherently control a large defect ensemble, suitable for sensing. We apply our methods experimentally to a spin ensemble of up to 4 $\times$ 10$^9$ nitrogen vacancy (NV) centers in diamond. By considering the physics of the system and explicitly including the hyperfine interaction in the optimization, we design shaped microwave control pulses that can outperform  conventional ($\pi$-) pulses when applied to sensing of temperature or magnetic field, with a potential sensitivity improvement between 11 and 78\%. Through dynamical modelling of the behaviour of the ensemble, we shed light on the physical behaviour of the ensemble system and propose new routes for further improvement.
\end{abstract}
\maketitle

\section{Introduction}

Solid state defects are a promising platform for quantum sensing, where purely quantum mechanical properties such as superposition and entanglement can be utilized to overcome classical limitations.\cite{Schroder2016, Aharonovich2016}. Particularly in semiconductors, where they can be controllably created and manipulated, solid state defects can host quantum states that are both long-lived and sensitive to the local environment in discrete energy levels within the bandgap. A typical and extensively used defect system is the nitrogen-vacancy (NV) center in diamond. This consists of a substitutional nitrogen atom and an adjacent lattice vacancy, having discrete electronic and nuclear spin states with long coherence times up to room temperature.\cite{Loubser1978} The optical properties of the negatively-charged NV center (NV$^{-}$) are highly sensitive to a range of parameters including magnetic field\cite{Taylor2008, Wang2012, Farfurnik2016, Farfurnik2018, Genov2019, Jelezko2004}, electric field\cite{Dolde2011, Wang2012}, temperature\cite{Neumann2013, Delord2017} and pressure (strain).\cite{Doherty2014} Applications include sensing using a scanning diamond tip\cite{Gross2017, Thiel2019}, nanoscale nuclear magnetic resonance (NMR)/ electron spin resonance (ESR)\cite{Staudacher2013, Lov2016} and in biophysics,\cite{Schirhagl2014, LeSage2013, Barry2016, Gorrini2018} where robustness and high biocompatibility of diamond makes it an ideal platform for sensing, even within biological samples.\cite{Kucsko2013, Fujiwara2020}

\begin{figure}[htbp]
\includegraphics[width=8.6cm]{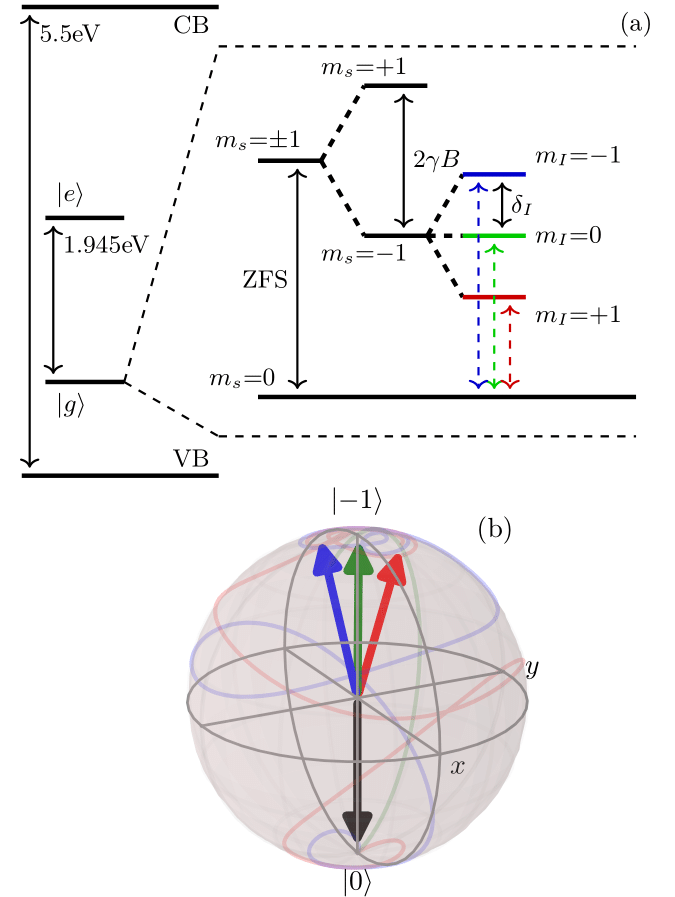}
\caption[]{Color online. (a) Simplified level diagram for a single NV$^{-}$ center within the diamond bandgap, with the ground state levels shown in detail.  At zero magnetic field there is a splitting  of \SI{2.87}{GHz} (ZFS) between the $m_{s}$=0 and $m_{s}$=$\pm$1 states. At finite field $B$, the Zeeman effect shifts the $m_{s}$=$\pm$1 states in energy by $\gamma B$. The $m_{s}$=$\pm$1 states are further split into 3 hyperfine levels ($m_l$=0,$\pm$1) separated by $\delta_{l}=\SI{2.16}{MHz}$.  (b) Bloch sphere representation depicting this $m_{s}=\{0,-1\}$ two-level system and the time evolution and result vector for a shaped optimal microwave pulse applied to the initial  ground state ($\ket{0}$, black arrow). Here we show the time evolution for each of the hyperfine resonances $m_{l}$. 
}
\label{fig:NV_levels}
\end{figure}

The level structure of the NV$^{-}$, illustrated schematically in Fig.~\ref{fig:NV_levels}(a) consists of spin triplet ground and excited states and metastable spin singlet states.\cite{Webb2020,Farfurnik2016, Rembold2020, Jelezko2004, Jelezko2006}. When green laser light is absorbed by an NV in $m_s$=0, red fluorescence is emitted from decay back into the triplet ground state. However, when absorbed in the spin-split $m_s$=$\pm$1, decay back to $m_s$=0 may occur through singlet shelving states, via nonradiative and infrared emission. The populations of $m_s$=0 and $\pm$1 can be controlled by applying resonant microwaves ($f=\SI{2.87}{GHz}$ in the absence of an external bias magnetic field). This results in a detectable decrease in red fluorescence output on resonance, with contrast $C$ of 1-\SI{2}{\%} for a large ensemble of defects and up to \SI{30}{\%} for a single NV\cite{Jelezko2006}. The $m_s$=$\pm$1 states can be split in energy e.g. via the Zeeman effect by an external magnetic field, giving rise to multiple spectral features including additional sub-features due to hyperfine splitting introduced by coupling to the nuclear spin of the \ce{^{14}N} or \ce{^{15}N} impurity atom\cite{Wojciechowski2018}. By sweeping microwave frequency, these resonances can be identified by the drop in fluorescence output, a process termed optically detected magnetic resonance (ODMR) spectroscopy. By fixing the microwave drive frequency on or close to a resonance, any frequency shift resulting from the level shift of $m_s$=$\pm$1 by magnetic field, electric field or local temperature can be detected.

Sensing using NV centers can be performed by a simple continuous wave (CW) method, maintaining a constant intensity of microwave and laser irradiation\cite{Rembold2020,Fescenko2020}. Alternatively, laser and microwave pulses can be used to control and read the ensemble\cite{Alkahtani2019, Jelezko2006}. This relies on the NV behaving as a two-level quantized system\cite{Dreau2011}, with one (bright) maximally fluorescent state, $\ket{0}$, and one (dark) state with reduced fluorescence under illumination with green light, $\ket{\pm 1}$. For a single NV, these correspond to the electron spin states $m_{s}$=0 and $m_{s}$=$\pm$1 respectively. Rabi oscillations can be observed in $C$ on application of a microwave field resonant with the ground state splitting. This allows coherent control using discrete laser and microwave pulses, offering improvement over  CW methods through reduction in the power broadening of the resonance linewidths. Techniques such as Ramsey interferometry \cite{Barry2019, Arai2018} and Hahn echo-type sequences have been demonstrated\cite{Nobauer2015a, Wolf2015}, realizing single molecule sensitivity in nanoscale diamond NMR experiments.\cite{Suter2017, Muller2014, Budker2019}

Pulsed schemes are used extensively for quantum sensing measurements using single or few-NV centers often in a confocal microscopy setup.\cite{Gruber1997,Balasubramanian2009,Balasubramanian2019} However, as extensive nuclear magnetic- and electron spin- resonance experiments have shown, a macroscopic ensemble of many billions of electron or nuclear spins in a larger volume can also be manipulated by microwave pulse sequences in the same manner.\cite{Charnock2001} From a quantum sensing perspective, large ensembles are desirable for imaging applications,\cite{LeSage2013} for vector sensing,\cite{Schloss2018} or to maximize bulk sensitivity where spatial resolution is not required, since the shot noise-limited sensitivity scales as $1 / \sqrt{N}$, with $N$ the number of defect centers.\cite{Taylor2008} Compared to single NV readout, ensemble NV sensing with flat (fixed amplitude and phase) microwave pulses suffers from nonuniform pulse operation. Inhomogeneous broadening due to strain and bias field gradients spreads the distribution of resonance frequencies of the NV centers, detuning many from a central drive frequency. In addition, the near-field microwave drive can vary in central resonance frequency, power, and phase across the ensemble, depending on the antenna and microwave coupling to the diamond.\cite{Horsley2018} This makes control of a large ensemble challenging. 

To overcome these issues, several approaches have been investigated for higher frequency AC sensing ($>\SI{10}{kHz}$), beginning with dynamical decoupling,\cite{Cai2012,deLange2010,Wolf2015,Farfurnik2018} and with further correction using e.g. adiabatic chirped pulses.\cite{Genov2019}. These are however unsuitable for applications that require DC to low frequency sensing, particularly for applications in biosensing\cite{Barry2016,Huotari2000,Kluess2010,Muniz2012}. An alternative in this frequency range is to deliver shaped microwave pulses (varying phase and amplitude), in order to boost fidelity in a Ramsey or pulsed ODMR\cite{Dreau2011} scheme. Such pulses can be designed using optimal control methods.\cite{Nobauer2015a, Nobauer2015, Bartels2013, Khaneja2005, Brif2010} Optimal pulses have been used with small ensembles of NV centers for Hahn-echo\cite{Nobauer2015a, Nobauer2015, Bjorn2015} or Carr-Purcell sequences,\cite{Hernandez-Gomez2019} to improve the robustness and temperature sensitivity of the D-Ramsey scheme,\cite{Konzelmann2018} to extend the coherence time of an NV,\cite{Dong2019} and to improve the accuracy of entanglement operations\cite{Dolde2014}. 

In this work, we demonstrate the use of shaped microwave pulses produced by optimal control methods combined with Floquet theory that can deliver improved coherent control over a large solid state defect ensemble of diamond NV centers.  We show improved ODMR contrast and therefore potentially higher sensitivity when compared to a conventional flat $\pi$-pulse sensing scheme. Our scheme is widely adaptable to a range of solid state systems where a two-level quantum system can be realized, although we specifically test our methods using an NV ensemble in diamond. We achieve our improvement through a full consideration of the physics of the system, including the hyperfine interaction with the nuclear spin of the subsitutional nitrogen in the NV center (both \ce{^{14}N} and \ce{^{15}N}). We model ensemble behaviour to further understand the physics of the system, in particular to explain the dynamics when a readout laser pulse is applied. This is also to uncover new routes for improvement for quantum sensing. We demonstrate our methods experimentally in off-the-shelf, standard grade material without significant processing or fabrication. Furthermore, we demonstrate operation at low Rabi frequencies, typical of those achievable using low-power microwave amplification e.g. in a portable sensor device.\cite{Webb2019}  

The paper is structured as follows. In Section~\ref{section:1} we outline the basic methodology we use to construct and generate our shaped microwave pulses using optimal control theory, including our derivation for explicitly including the hyperfine interaction in the optimization algorithm. We describe a number of key control parameters, the limits of which we discuss in Section~\ref{section:2}. In Section~\ref{section:3} we describe in detail our experimental setup and methodology and in Sections~\ref{section:4} and \ref{section:5} we demonstrate the use of optimized shaped pulses for ODMR spectroscopy, compare to a conventional $\pi$-pulse scheme using a flat microwave pulse, and analyze and discuss the optical behaviour and how this relates to the physical dynamics of the NV ensemble. 

\section{Methods} 
\subsection{Optimal Control} 
\label{section:1}

Our optimal control algorithm maximizes a functional that describes the desired transfer of one quantum state to another.\cite{Nobauer2015a, Nobauer2015, Khaneja2005, Poggiali2018, Rembold2020, Brif2010}. We define our state transfer functional as:
\begin{equation} \label{eq:stF}
F_\text{st} = \left| \left\langle \psi _f \left| \hat{U} ( t_p ) \right| \psi _i \right\rangle \right| ^2,
\end{equation}

where $\mathcal{F}_\text{st}$ is the fidelity, of value between 0 and 1, which describes how well the pulse transfers the quantum state of a solid state defect from an initial state $\left| \psi _i \right\rangle$ to a final state $\left| \psi _f \right\rangle$. A fidelity of 1 represents a complete transfer to the desired state. The influence of the pulse is described by the unitary time evolution operator $\hat{U} (t_p)$, where $t_p$ is the pulse duration. 

To represent the state transfer of an ensemble, we calculate ${F}_\text{st}$ for each member of a sample of defects with a specified range of frequency detuning $\hat{\Delta}$ and relative control amplitude $\hat{\alpha}$. These factors are set to be representative of the variation across a real ensemble. The relative control amplitudes $\alpha_i$ represent the drive field inhomogeneity across the ensemble, and each value is the ratio between the Rabi frequency at which a given single defect is driven (due to drive field inhomogeneity) and the Rabi frequency at which the pulse is designed to drive the defects. The values of $\alpha_i$ thus vary around unity across the representative sample. The relative control amplitude only relates to the changes in Rabi frequency caused by drive field inhomogeneity and does not include the effects of frequency detuning on the Rabi frequency. These effects are included in the optimization separately via the $\Delta_i$ values, which represent the inhomogenous broadening. We thus assign each defect in the representative sample a value of $\alpha_i$ and $\Delta_i$ within the specified range $\hat{\Delta}$ and $\hat{\alpha}$ and seek a pulse that maximizes the average fidelity of the entire representative ensemble. Using this model assumes that interaction between defects is minimal, such that each defect can act as as single, isolated quantum defect in the material. 

\begin{figure}[h]
    \centering
    \includegraphics[width=8.6cm]{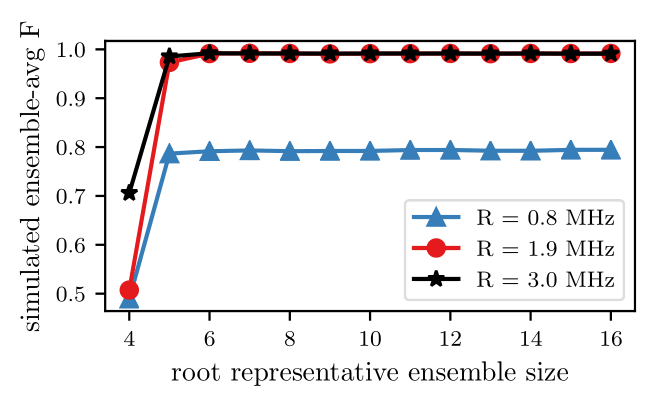}
    \caption{Color online. The simulated weighted average fidelity of optimal control pulses optimized with different ensemble sizes as a function of representative ensemble size for three values of the maximum allowed Rabi frequency $R_\text{lim}$. The pulses were optimized for $\hat{\Delta} = \pm \SI{1}{MHz}$ detuning, $\hat{\alpha} = 1\pm \SI{10}{\%}$ amplitude variations and a duration of $t_p =\SI{1.85}{\micro\second}$ with the indicated values of $R=R_\text{lim}$.}
    \label{fig:ensemble_size}
\end{figure}

We assume our detunings $\Delta _i$ follow a Gaussian distribution centered at zero. The full width at half maximum (FWHM) of this Gaussian distribution is set equal to half of the width of the considered detuning range $\hat{\Delta}$. The $\alpha_i$ values are assumed to follow a flat distribution. The weight of each defect in the representative ensemble is thus equal to the weight of its $\Delta_i$ value. These are normalized such that the sum of the weights of all defects in the representative ensemble is equal to 1. We therefore also use a weighted average of the fidelity. For numerical optimization, we use throughout this work a representative ensemble of size 12x12 (12 values to cover the ranges $\hat{\Delta}$ and $\hat{\alpha}$, respectively). This was based on a series of simulations of the performance of pulses transferring state $\ket{0}$ to $\ket{-1}$ (Fig.~\ref{fig:NV_levels}(b)) optimized using different representative ensembles. As shown in Fig.~\ref{fig:ensemble_size}, 12x12 more than ensures convergence of the fidelity, while minimizing computational time.

\begin{figure}[h]
    \centering
    \includegraphics[width=8.6cm]{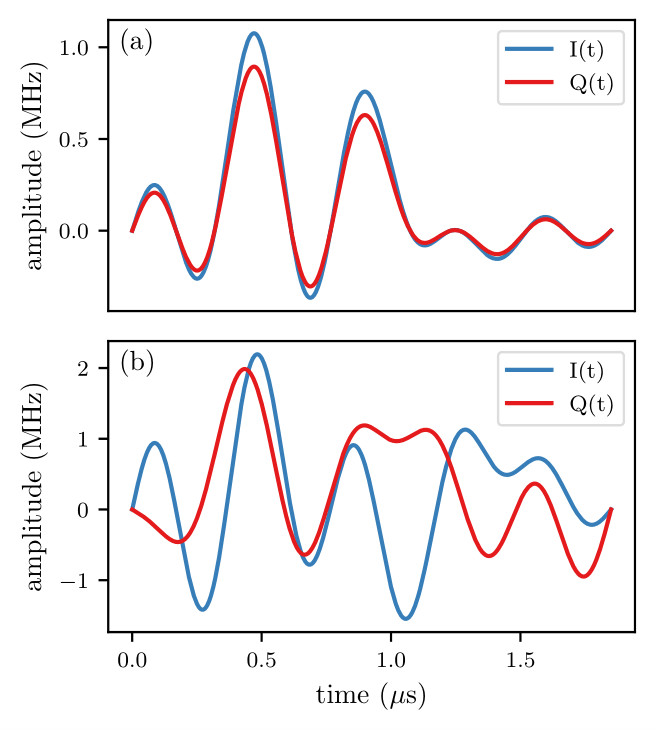}
    \caption{Color online. Plot of $I(t)$ and $Q(t)$ in units of Rabi frequency for two optimal control pulses that were optimized by including state transfer using all three hyperfine levels. The pulses were optimized for $\hat{\Delta} =\pm \SI{1}{MHz}$ detuning and $\hat{\alpha} = 1\pm \SI{10}{\%}$ amplitude variations with a duration of $t_p = \SI{1.85}{\micro\second}$ and a maximum allowed Rabi frequency of (a) \SI{1.4}{MHz} and (b) \SI{3.0}{MHz}.}
    \label{fig:IQ}
\end{figure}

For the design of our shaped microwave pulses, we use smooth optimal control. Here we choose a basis of periodic functions with the same periodicity $T$ and discretized frequency components, resulting in the shaped pulses becoming smooth in time.\cite{Bartels2013} In this work, we use a basis of sine functions with a fundamental frequency determined by the pulse duration $t_p$.\cite{Nobauer2015, Nobauer2015a} Smooth optimal control has experimental advantages over alternatives such as gradient ascent pulse engineering (GRAPE)\cite{Skinner2003} in that the bandwidth and the individual frequency components are known in advance and the number of high frequency components in the pulse Fourier spectrum is reduced, making modulation in experiments less technically demanding.\cite{Bartels2013} Our smooth optimal control pulse has the general form:

\begin{equation} \label{eq:optS}
S(t) = I(t) \cos \left( \omega _{D} t \right) + Q(t) \sin \left( \omega _{D} t \right),
\end{equation}
where
\begin{equation} \label{eq:optIQ}
I(t) = \sum\limits_{j=1}^{N_f} 2 a_{jx} \sin \left( j \Omega _f t \right), \: Q(t) = \sum\limits_{j=1}^{N_f} 2 a_{jy} \sin \left( j \Omega _f t \right).
\end{equation}
Here, $\omega _{D}$ is the central driving frequency, $\Omega _f = 2 \pi / (2 t_p )$ is the fundamental frequency, $N_f$ is the number of frequency components and the real $a_{jk}$-values are control amplitudes. The bandwidth of such a pulse is then $N_f \Omega$. The fundamental frequency is not related to the Rabi frequency and purely serves to enforce the desired periodicity of $T=2 t_p$. The $a_{jk}$-values are, however, defined in units of Rabi frequency. As an example, Fig.~\ref{fig:IQ} shows the in-phase and quadrature components $I(t)$ and $Q(t)$ used to modulate the microwave carrier for two of the specific pulses that we designed. In our experiments, the microwave carrier has a frequency $\omega_{D} \approx \SI{2.8}{GHz}$ corresponding to the splitting between the $m_s$=0 and $m_s$=-1 levels of the NV center ground state with an applied bias magnetic field.  

It has been previously shown\cite{Bjorn2015} that the performance of smooth optimal control pulses improves with increasing $N_f$ until it saturates for $N_f \geq 7$. We use $N_f=10$ for all of our pulses to ensure that we are in the saturated regime. This yields 20 different control amplitudes $a_{jk}$ per shaped pulse, and these are the parameters that are optimized by the control algorithm. The optimization is carried out iteratively by stepping along the gradient of the fidelity with respect to the control amplitudes with a step size $\beta$. Starting with initial control amplitudes $a_{jk}$, we compute the resulting $\hat{U}(t_p)$, $\mathcal{F}_\text{st}$ and $\frac{\partial \mathcal{F}_\text{st}}{\partial a_{jk}}$, before updating the control amplitudes by adding $\beta \frac{\partial \mathcal{F}_\text{st}}{\partial a_{jk}}$. This process is then repeated until $\mathcal{F}_\text{st}$ converges. The choice of time-periodic basis functions yields a time-periodic Hamiltonian that can be solved using Floquet theory.\cite{Goelman1989, Skinner2010, Bartels2013}

In this work, we extend previous methods to include the effects of hyperfine splitting during the optimization. Although we specifically calculate for diamond NV centers here, this method is generally adaptable and applicable to any such splitting for a defect ensemble. The goal is to create a shaped pulse that performs the state transfer $\ket{0}$ to $\ket{-1}$ simultaneously and with as high fidelity as possible for each of the $m_l$ hyperfine levels. For an NV center ensemble, this results in a higher ODMR contrast than would be otherwise achievable by acting on only one $m_l$. This is analogous to continuous wave methods driving multiple hyperfine lines previously described in the literature.\cite{El-Ella2017} By doing this in the pulsed domain, we seek to achieve similar benefits, but without the negative effects of power broadening of the resonance linewidths.  In order to explicitly account for the hyperfine splitting, it is necessary to modify the expression for the Fourier components of the Hamiltonian that make up the Floquet matrix. The Fourier components of the Hamiltonian are generally defined as

\begin{equation} \label{eq:Hn}
    \hat{\mathcal{H}}_n = \frac{1}{T} \int_{0}^{T} \exp \left( -i n \Omega _f t \right) \hat{\mathcal{H}}(t) \, \mathrm{d} t,
\end{equation}

where $T = 2 t_p$ is the periodicity of the Hamiltonian and $\hat{\mathcal{H}} (t)$ is the time-domain Hamiltonian that describes the system to be optimized. The nitrogen in an NV can be either \ce{^{14}N} with $I=1$ (highest natural abundance) or \ce{^{15}N} with $I=1/2$, yielding either three or two hyperfine levels, respectively, as illustrated in Fig.~\ref{fig:NV_levels}(a). We assume hyperfine interaction between the \ce{^{14}N} nuclear spins and the NV electron spins in the ensemble so that three hyperfine states are possible. The nuclear spins are assumed to be in a thermal state such that all $m_l$ hyperfine states are equally represented in the ensemble. The ODMR spectrum then contains three resonances separated by $\delta_{l}=\SI{2.16}{MHz}$, corresponding to the three hyperfine resonances $m_l$={-1,0,1}. We also assume that the different NV electron spins do not interact and that the $m_s$=$\pm$1 states are clearly split by a static magnetic bias field. A single set of three NV centers that each correspond to one of the hyperfine transitions can then be reasonably approximated as three independent two-level systems. The drift Hamiltonian thus has the form

\begin{equation}
    \hat{\mathcal{H}}_0 = \sum\limits_{k=1}^{3} \frac{\omega_{0,k}}{2} \sigma_{z,k},
\end{equation}

where $\omega_{0,k}$ is the transition frequency of hyperfine transition $k$, and $\sigma_{z,k}$ is a Pauli spin-$z$ matrix that is specific to transition $k$. Note that the above expression applies to any two-level defect with three equidistant hyperfine resonances that fulfills the underlying assumptions. The transition frequencies are related via $\omega_{0,1} = \omega_{0,2} - \delta_{l}$ and $\omega_{0,3} = \omega_{0,2} + \delta_{l}$. Given that the states of the three two-level systems can be completely described by a single vector of length 6, the $\sigma_{z,k}$-matrices can also be represented by 6-by-6 matrices. (See Appendix~\ref{App:spin}). The same is true of the $\sigma_{x,k}$- and $\sigma_{y,k}$-matrices. The control Hamiltonian describes the interaction between the control pulse of the form given in Eq.~(\ref{eq:optS}) and the three allowed transitions.

Assuming the control field is linearly polarized in the $x$-direction, which is perpendicular to the defect axis, the control Hamiltonian can be written in the form:
\begin{equation}
    \hat{\mathcal{H}}_c = \sum\limits_{k=1}^{3} \sigma_{x, k} \left[ I(t) \cos \left( \omega _{D} t \right) + Q(t) \sin \left( \omega _{D} t \right) \right],
\end{equation}
and the total Hamiltonian thus has the form
\begin{multline} \label{eq:Hinit}
    \hat{\mathcal{H}} (t) = \sum\limits_{k=1}^{3} \left( \frac{\omega_{0,k}}{2} \sigma_{z,k} \right. \\ \left. + \sigma_{x, k} \left[ I(t) \cos \left( \omega _{D} t \right) + Q(t) \sin \left( \omega _{D} t \right) \right] \right).
\end{multline}
We can simplify the rest of the calculations by working in a rotating frame given by the unitary rotation operator
\begin{equation}
    \hat{R} = \exp \left( \sum\limits_{k=1}^{3} i \omega_{D} t \sigma_{z, k} / 2 \right),
\end{equation}

which will commute with every term in $\hat{\mathcal{H}}_c$ except for $\sigma_{x,k}$. More precisely, $\left[\sigma_{z,k}, \sigma_{x,k'} \right] = 2 i \sigma_{y, k} \delta_{k,k'}$ and $\left[\sigma_{z,k}, \sigma_{y,k'} \right] = - 2 i \sigma_{x, k} \delta_{k,k'}$.

The Baker-Campbell-Hausdorff lemma thus allows us to write
\begin{multline}
    \hat{R} \hat{\mathcal{H}}_c \hat{R}^{\dagger} =  \sum\limits_{k=1}^{3} \left( \sigma_{x,k} \cos \left( \omega_{D} t \right) + \sigma_{y,k} \sin \left( \omega_{D} t \right) \right) \\ \times \left[ I(t) \cos \left( \omega _{D} t \right) + Q(t) \sin \left( \omega _{D} t \right) \right].
\end{multline}
Using this expression and defining the detuning, $\Delta = \omega_{0, 2} - \omega_{D}$, as the difference between the transition frequency of the central hyperfine transition, $\omega_{0, 2}$, and the central driving frequency, $\omega_{D}$, we obtain the expression
\begin{multline}
    \hat{\mathcal{H}}' = \sum\limits_{k=1}^{3} \left( \frac{\Delta + w_k \delta_{l}}{2} \sigma_{z,k} \right. \\ \left. + \left( \sigma_{x,k} \cos \left( \omega_{D} t \right) + \sigma_{y,k} \sin \left( \omega_{D} t \right) \right) \right. \\ \left.
    \times \left[ I(t) \cos \left( \omega _{D} t \right) + Q(t) \sin \left( \omega _{D} t \right) \right] \right),
\end{multline}
where $w_1 = -1$, $w_2 = 0$ and $w_3 = 1$. Expanding by using trigonometric relations, the above expression can be simplified by using the rotating wave approximation to eliminate the fast-oscillating terms

\begin{equation} \label{eq:Hsim}
    \hat{\mathcal{H}}' = \sum\limits_{k=1}^{3} \left( \frac{\Delta + w_k \delta_{l}}{2} \sigma_{z,k} + \frac{I(t)}{2} \sigma_{x,k} + \frac{Q(t)}{2} \sigma_{y,k} \right).
\end{equation}
Combining Eq.~(\ref{eq:Hsim}) with Eq.~(\ref{eq:optIQ}) and inserted into Eq.~(\ref{eq:Hn}), the Fourier components of the Hamiltonian become

\begin{multline}
    \hat{\mathcal{H}}_n = \sum\limits_{k=1}^{3} \frac{1}{T} \int\limits_{0}^{T} \exp \left( i n \Omega t \right) \left( \frac{\Delta + w_k \delta_{l}}{2} \sigma_{z,k} \right. \\ \left. + \sum\limits_{j=1}^{N_f} \left[ a_{jx} \sigma_{x,k} + a_{jy} \sigma_{y,k} \right] \sin \left( j \Omega t \right) \right).
\end{multline}
The above expression can be further simplified by using the exponential form of a sine and the integral form of a Kronecker delta. Doing so yields the final expression for the Fourier components of the Hamiltonian when the effects of hyperfine splitting are taken into account
\begin{multline} \label{eq:Hnfinal}
    \hat{\mathcal{H}}_n = \sum\limits_{k=1}^{3} \left( \frac{\Delta + w_k \delta_{l}}{2} \sigma_{z,k} \delta_{n,0} \right. \\ \left. + \sum\limits_{j=1}^{N_f} \frac{1}{2i} \left[ a_{jx} \sigma_{x,k} + a_{jy} \sigma_{y,k} \right] \cdot \left[ \delta_{n, j} - \delta_{-n, j} \right] \right)
\end{multline}

We use Eq.~(\ref{eq:Hnfinal}) in the construction of the Floquet matrix for the computation of $\hat{U} (t_p)$ and $\frac{\partial \mathcal{F}}{\partial a_{jk}}$ as part of the update step of the optimal control algorithm. We include the corresponding derivation for two hyperfine levels (\ce{^{15}N} for NV centers) in the Supplementary Information. Control amplitude variations are included by multiplying the control amplitudes $a_{jx}$, $a_{jy}$ by the $\alpha_i$-value for the given defect in the representative ensemble. 

In order to ensure the optimization of our control amplitudes converges while remaining within experimentally achievable limits, we include a penalty functional 
\begin{equation}
\mathcal{F}_\text{pen} = - p t_p \sum\limits_{j,k} a_{jk}^2
\end{equation}

in our algorithm, applied at each update step. The penalty functional includes a specified penalty constant $p> 0$ and scales with the control amplitudes. We optimize using the gradient of the sum of the penalty functional and the state transfer fidelity $\mathcal{F}_\text{tot}$=$\mathcal{F}_\text{pen}$+$\mathcal{F}_\text{st}$. After each update step, the maximum amplitude of the optimal control pulse is computed in units of Rabi frequency, and if it exceeds the maximum allowed Rabi frequency $R_\text{lim}$, the penalty constant is increased by a step size $\Delta p$. If the maximum amplitude of the optimal control pulse does not exceed $R_\text{lim}$, the penalty constant is reduced by $\Delta p$. $R_\text{lim}$ is one of the inputs to the algorithm and is limited by the maximum achievable experimental Rabi frequency $R_\text{max}$. This method also prevents the algorithm remaining in local maxima compared to optimizing without a penalty functional.

\begin{figure*}
    \centering
    \includegraphics[width=\textwidth]{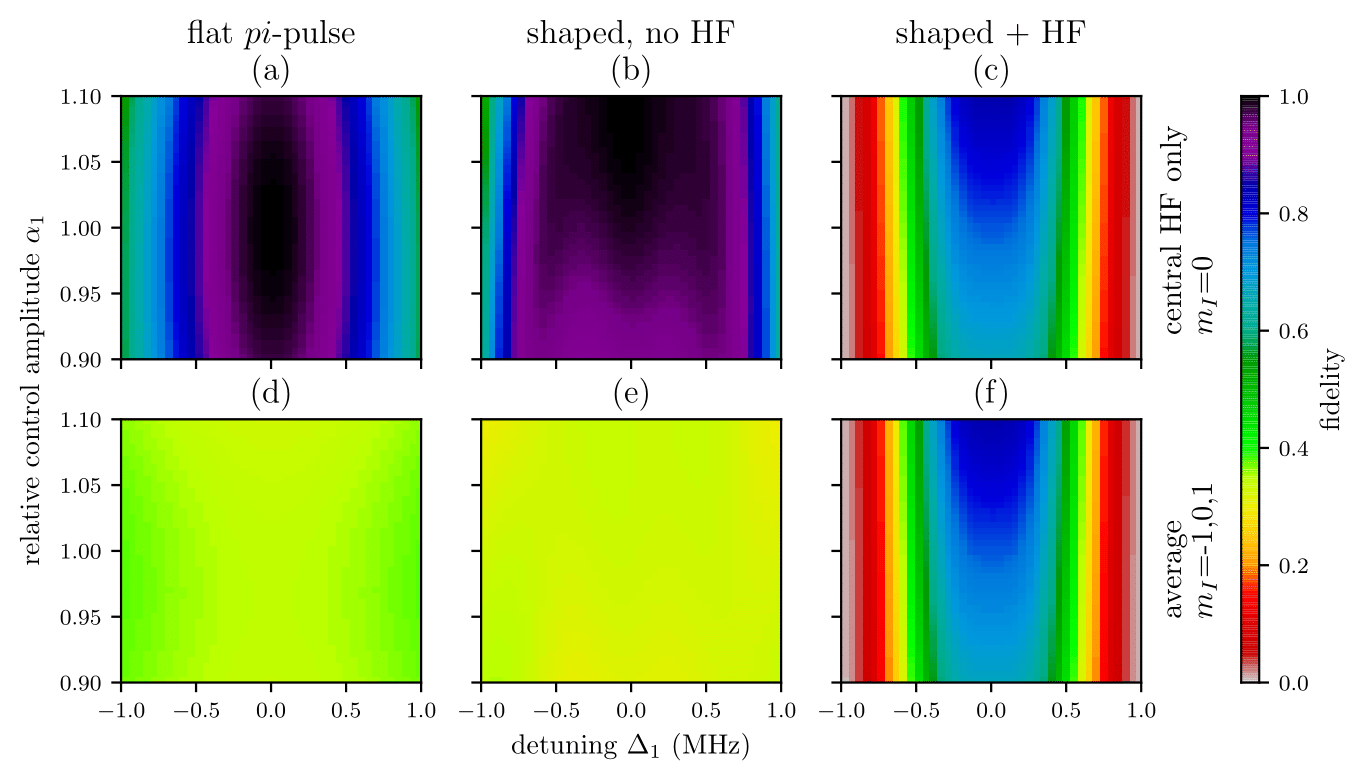}
    \caption{Color online. Simulated maps of the state transfer fidelity (Eq.~(\ref{eq:stF})) from $\left|0 \right\rangle$ to $\left|1 \right\rangle$ for a single NV (i=1) subject to (a,d) a flat ($\pi$)-pulse, (b, e) an optimized shaped pulse and (c, f) a shaped pulse optimized while taking the fidelity of state transfer taking all three hyperfine levels into account. The top plots (a, b, c) show the fidelity of the transfer experienced by the central hyperfine transition while the bottom plots (d, e, f) show the average of the transfer fidelities for each of the three hyperfine levels. Each point in the top plots is the fidelity for a single NV electron spin with the given values of $\alpha_1$ and $\Delta_1$. Each point in the bottom plots is the average of the fidelities for three NV electron spins with the given value of $\alpha_1$ and transition frequencies detuned by $\Delta_1$, $\Delta_1 + \delta_{l}$ and $\Delta_1 - \delta_{l}$, respectively, from the driving frequency. The flat pulse has a Rabi frequency of \SI{1.4}{MHz}, and the optimal control pulses were both optimized using $\hat{\Delta} = \pm \SI{1}{MHz}$ detuning, $\hat{\alpha} = 1\pm \SI{10}{\%}$ amplitude variation, $R_\text{lim}$= \SI{1.4}{MHz} and a pulse duration $t_p = \SI{1.85}{\micro\second}$.}
				\label{fig:Fmap}
\end{figure*}

As a demonstration of the effect of explicitly including all three hyperfine levels in the optimization, Fig.~\ref{fig:Fmap} shows a series of simulated fidelity maps for a single NV subject to a flat $\pi$-pulse and optimal control pulses with and without including the hyperfine components. The fidelity of a $\ket{0}$ to $\ket{-1}$ state transfer is directly proportional to the resulting ODMR contrast $C$ since the contrast will be maximal when all NV electron spins are in the $\left|-1\right\rangle$ state and minimal in the $\left|0\right\rangle$ state. All three pulses are in the regime $R_\text{lim}<\delta_{l}$. It is clear that the regular optimal control pulse has superior performance for a single hyperfine resonance. However, when considering the average of all three, the shaped pulse optimized while taking the effects of hyperfine splitting into account is significantly better, albeit within a narrower range of detuning. Fig.~\ref{fig:Fmap}(f) indicates that the optimal control pulse including the hyperfine splitting in the optimization is capable of simultaneously performing state transfer using all three hyperfine levels with high fidelity. The narrow range of high fidelity dropping rapidly with detuning indicates that the optimal pulse will yield high contrast when applied with drive frequency $\omega_D$ close to any one of the three hyperfine resonances and low contrast when applied off-resonance. This behavior naturally translates to a high contrast and narrow resonance linewidth and thus higher sensitivity to magnetic field. As can be seen in Fig.~\ref{fig:Fmap}(c,f), as $\alpha_1$ is increased, the $\ket{0}$ to $\ket{-1}$ fidelity (i.e., ODMR contrast) further improves in the narrow range of high fidelity without significantly broadening the range of high fidelity. We therefore experimentally apply our optimal control pulses at applied microwave power equivalent to a higher maximum Rabi frequency than we use for optimization, empirically chosen to maximize the slope. 

\subsection{Optimization Details}
\label{section:2}

All of our pulses were made using an initial value of the penalty constant $p=1$ and $\Delta p = 0.05$. They were optimized to perform a state transfer from $\left|0 \right\rangle$ to $\left|-1 \right\rangle$. We used 150 update steps for all of the optimizations, as this was found to be sufficient to achieve convergence of $\mathcal{F}_\text{st}$. For the first 51 steps, the step size along the gradient was kept constant at $\beta = 0.007$ and for the remaining steps, the optimal step size was determined using a line search. This was done to speed up the optimization without compromising the quality of the resulting optimal control pulses. We designed pulses using different values of $R_\text{lim}$, $t_p$ and the ranges $\hat{\Delta}$ and $\hat{\alpha}$ and tested them experimentally. We determined the maximum achievable experimental Rabi frequency, i.e. the upper limit on the maximum allowed Rabi frequency $R_\text{lim} \leq R_\text{max}=\SI{3.2}{MHz}$ through prior experimental measurements using flat pulses on the same diamond NV ensemble. Based on this, we defined a range of $R_\text{lim}$ to generate testable optimized shaped pulses for as between $R_\text{lim}=\SI{0.8}{MHz}$ and $R_\text{lim}=\SI{3.2}{MHz}$. The minimum value of $t_{p}$ necessary to achieve improvements over a comparable flat pulse was limited by the need to apply sufficient power to perform the desired state transfer. We set the lower limit of $t_p$ to be at least twice the duration of a flat $\pi$-pulse with Rabi frequency equal to $R_\text{lim}$. The maximum value of $t_p$ was limited by the $T_{2}$ coherence time of a single NV. Based on this, we defined a range of $t_p$ to generate testable optimized shaped pulses for as between $t_p=\SI{1.0}{\micro\second}$ and $t_p=\SI{5.0}{\micro\second}$. 

Although the possible values of detuning $\Delta_i$ are in principle not limited, higher Rabi frequencies are required to compensate for higher levels of inhomogenous broadening. Based on the considered values of $R_\text{lim}$, we therefore used $\hat{\Delta}$ up to $\pm \SI{2}{MHz}$. The possible values of $\alpha_i$ are similarly not limited in principle, but higher Rabi frequencies are required to compensate for higher levels of drive field inhomogeneity. We therefore chose to optimize up to $\hat{\alpha} = 1 \pm 0.2$ relative control amplitude range.

Our initial  $a_{jk}$-values were set using pseudorandom values within a range sufficient to yield a maximum Rabi frequency of the corresponding initial pulse greater than $R_\text{lim}$. This was done in order to ensure that the optimization algorithm approached the region of allowed pulses from the outside, so that pulses utilizing $R_\text{lim}$ were considered. For this work, the initial Rabi frequency was 2.8 times greater than the maximum allowed Rabi frequency.

\subsection{Experimental Setup}
\label{section:3}
\begin{figure}
  \includegraphics[width=8.6cm]{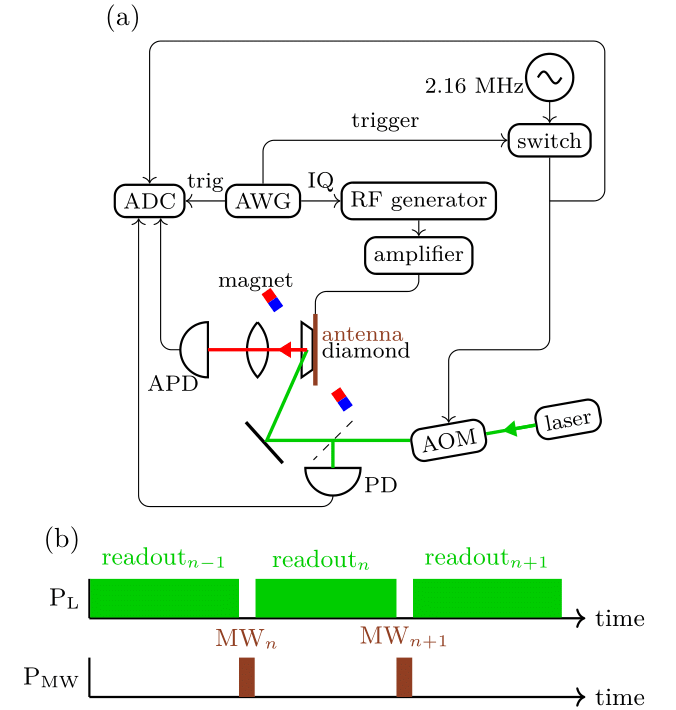}
  \caption{Color online. (a) Schematic of our experimental setup. The pump laser is modulated by the AOM, at $\SI{2.6}{MHz}$ and controlled by the AWG. Microwave pulses are delivered to the diamond using a near-field antenna. The AWG provides IQ modulation to the signal generator to create the required control pulses. An ADC, synchronized with the AWG, digitizes the analog AOM modulation signal, the signal from the APD collecting the diamond fluorescence $V_{\text{fl}}$ and the signal from an amplified photodetector that collects a small amount of the pump laser $V_{\text{ref}}$. (b) Pulsed ODMR sequence as applied in our measurements, showing the repeating sequence of pump laser pulses $P_l$ and microwave pulses $P_\text{MW}$. This sequence is repeated continuously by the AWG.}
  \label{fig:setup}
\end{figure}

A schematic of our experimental setup is shown in Fig.~\ref{fig:setup}(a). We used an off-the-shelf, optical-grade diamond (Element 6) with $\sim\SI{0.5}{ppb}$ NV$^-$ concentration, of dimensions 5x5x\SI{1.2}{mm^3}. For this diamond, we measured a $T_2^*$-limited linewidth of \SI{0.75}{MHz} and determined $T_1$, $T_2$ and $T_{2}^{*}$ times as \SI{7.1}{ms}, \SI{7.0}{\micro\second} and \SI{0.44}{\micro\second} respectively, with a maximum ensemble-averaged Rabi frequency of $R_\text{max}=\SI{3}{MHz}$ driven by our antenna (see Supplementary Information for details). A bias field of \SI{2.9}{\milli\tesla} aligned along the [111] crystallographic axis was applied by fixed permanent magnets, so as to split the $m_s$=$\pm$1 states. We addressed only the $m_s$=$0\xrightarrow{}m_s$=$-1$ transition to use an effective two-level system within the antenna's resonance.

The diamond was optically pumped using a \SI{532}{\nano\meter} diode-pumped solid state laser (DPSS, Cobolt Samba 1500). The linearly polarized beam was focused to a waist diameter of $~$\SI{120}{\micro\meter} before Brewster-angle refraction into the diamond to optically address (with at least $1/e^{2}$ the center intensity) an ensemble with a minimum estimated size of 
$\approx\SI{4d9}{}$ NV centers in a volume of $\approx\SI{0.04}{mm^3}$ based on the focused waist of the pump beam. The maximum pump laser power we delivered to the diamond was \SI{500}{mW}. This resulted in \SI{84}{\micro\watt} of red fluorescence escaping the front face of the diamond, of which we collected \SI{9.1}{\micro\watt} by using two condenser lenses (Thorlabs ACL25416U) to first collimate to pass through a low-pass filter (FEL0550) and then focus onto an avalanche photodiode (Thorlabs APD120A), producing an amplified analog voltage output $V_{\text{fl}}$ sampled by an analog-to-digital converter (ADC, Gage Octopus CS8300) at \SI{50}{MHz}. We optically modulated our pump laser using an acousto-optic modulator (AOM, Isomet 532C-4) at $f_\text{AOM}=\SI{2.6}{MHz}$, allowing us to perform software lock-in detection to minimize noise in the electronic readout. A fraction of the pump beam was also sampled by a second detector (Thorlabs PDA10A) to provide a reference, $V_{\text{ref}},$ for common-mode noise rejection.

We generated the microwave pulses necessary for implementation of the optimal control protocols using an arbitrary waveform generator (AWG, Tektronix 5000), in-phase/quadrature (IQ) modulating a Stanford SG394 RF signal generator. The microwave output was amplified (Mini-Circuits ZHL-16W-43-S+) and delivered to the diamond using a near-field antenna based on a square split-ring resonator design \cite{AhmadiThesis, GencResonator}. This antenna was designed for uniformity of near-field intensity in a 5x\SI{5}{mm^2} region centered on the diamond with a resonance at approximately \SI{2.8}{GHz}. Our AWG also controlled a switch (Minicircuits ZASWA-2-50DRA+) through which the AOM modulation drive was passed, allowing the pump beam incident on the diamond to be pulsed and modulated.

\subsection{Pulse Sequencing and Readout}

In our experimental setup, we measured contrast $C$, the change in fluorescence output as a result of a control pulse. We define $C$ as the change in fluorescence output in the initial period of a pump laser readout pulse after application of a preceding microwave pulse\cite{Dreau2011,Levine2019,Wolf2015}. $C$ was measured across an ODMR resonance feature by varying microwave drive frequency $\omega_D$. We measured this change in fluorescence signal $V_{\text{fl}}$ after application of either a shaped or flat microwave pulse, relative to the laser reference signal $V_{\text{ref}}$. We obtained $C$ by scaling the reference to the size of the fluorescence signal, subtracting the two, and integrating the resulting signal over a time window $t_w$=0.3-2.7 ms at the start of the laser pulse (see Supplementary Information for full details). This subtraction method allowed us to reject both DC and higher-frequency ($>$kHz) common-mode noise from the laser on the readout signal within the integration window. It also allowed us to measure a value for $C$ from every laser pulse (rather than measuring a reference with no microwave pulse on every other fluorescence readout), maximizing the bandwidth of our readout. From $C$ we also derived $C^{'}$ the change in contrast with microwave drive frequency. This quantity, the slope of the ODMR resonance, gives a measure of the strength of response and hence sensitivity to the local environment. 

Using the pulsed protocol shown in Fig.~\ref{fig:setup}(b), we first initialized the NV ensemble into the ground state using pump laser pulse $(n-1)$ of duration $t_l$. The pump laser was then blocked by the AOM during application of microwave control pulse $n$ of duration $t_p$. A subsequent laser pulse $n$ of the same duration $t_l$ was then applied and the state read out via diamond fluorescence emission This pulse also acted to reinitialize the system back into the $\ket{0}$ state, allowing the next $(n+1)$ pulses to read and initialise. This method enabled measurements using only a short repeating sequence in the AWG memory. 
We acquired data continuously for repeated sequence sets up to the memory limit of the ADC ($n$=110 pulses when using $t_l$=\SI{3}{ms}). Once this limit was reached, the data was transferred to computer memory and processed, averaging over all pulse sequence sets in the acquisition to reduce noise, and then integrating to obtain $C$.  

For direct comparison, we performed the same pulse sequence with the same readout methods for $C$ using both shaped microwave pulses and standard fixed amplitude and phase (flat) pulses. We used the same method for calculating $C$ throughout our measurements, to ensure accurate comparison between the different microwave pulses. In the latter case, we used pulses with a single microwave drive frequency of the form $\cos(\omega t)$ and three-frequency drive pulses of the form $\sum_{n\in \{ 0,\pm 1 \}} \cos((\omega + n \delta_{l})t + \phi_{n})$ to drive multiple hyperfine transitions.\cite{OptimisedFmod} The latter were generated using the AWG with randomized phases $\phi_{n}$ for each ADC acquisition to eliminate time-dependent artifacts.

\section{Results}

\subsection{Laser Pulse Duration}
\label{section:4}
Our previous measurements\cite{OptNVBio} demonstrated long optical reinitialization times, requiring many milliseconds on an approximately exponential decay with laser pulse duration to fully return the ensemble to the ground state. For the comparably sized ensemble in these experiments, we observed similar exponential behavior with a time constant of of $\approx$\SI{1.4}{ms}. 

\begin{figure}[htbp]
\includegraphics[width=8.6cm]{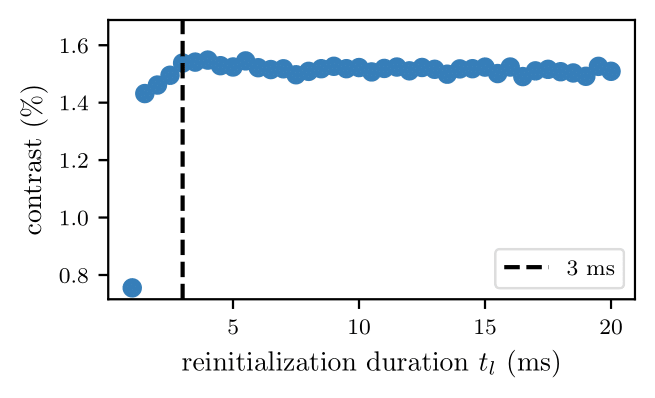}
\caption[]{Contrast as a function of laser pulse time $t_l$. Below 3 ms, $t_{l}$ is too short to sufficiently reinitialize the ensemble, leaving to a reduction in contrast $C$ with shorter readout/reinitilisation laser pulse length $t_l$.}
\label{fig:contrastvsinit}
\end{figure}

\begin{figure}[htbp]
\centerline{\includegraphics[width=8.6cm]{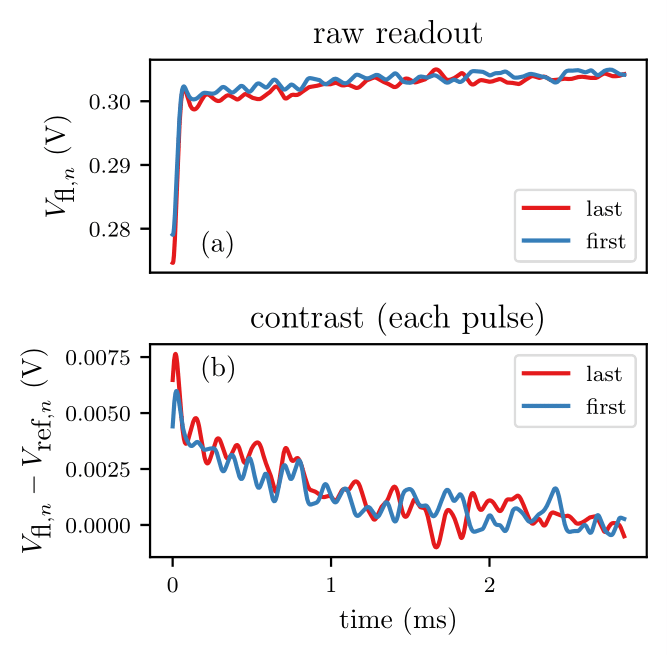}}
\caption[]{Raw fluorescence readout signal $V_{\text{fl},n}$ and relative contrast $C_{r,n}(t) = V_{\text{fl},n} - V_{\text{ref},n}$ for the first (n=1) and last (n=110) \SI{3}{ms} readout laser pulse in a single ADC acquisition of 110 readout sequences. No difference within the readout noise was observed at this readout duration, as would be expected from hysteresis effects arising from insufficient reinitialization of the ensemble.}
\label{fig:hyst}
\end{figure}

Waiting tens of milliseconds per readout would severely limit the number of pulses we could read and average in a single ADC acquisition and thus our contrast resolution. We therefore first performed experiments varying laser pulse duration to determine whether we could initialize and control the ensemble using shorter laser pulses without suffering hysteresis effects, either from incomplete initialization or reionization delay across the readout laser pulses.\cite{Giri2018, Manson2005, Aslam2013}

Fig.~\ref{fig:contrastvsinit} shows the contrast $C$ as a function of laser readout pulse duration $t_l < \SI{20}{ms}$ as measured using an optimal control pulse. We observed $C$ to be reduced for times shorter than $\approx\SI{3}{ms}$, indicative that an increasing number of NV centers in the sample were not fully reinitialized into the ground state. For $t_l$=3ms and above, we observed negligable hysteresis effects in the fluorescence readout. This is supported by Fig.~\ref{fig:hyst}, comparing the raw fluorescence readout and relative contrast calculated from the first and last individual readout pulses in a 110 pulse ADC acquisition using $t_l=\SI{3}{ms}$.

\begin{figure}[htbp]
\centerline{\includegraphics[width=8.6cm]{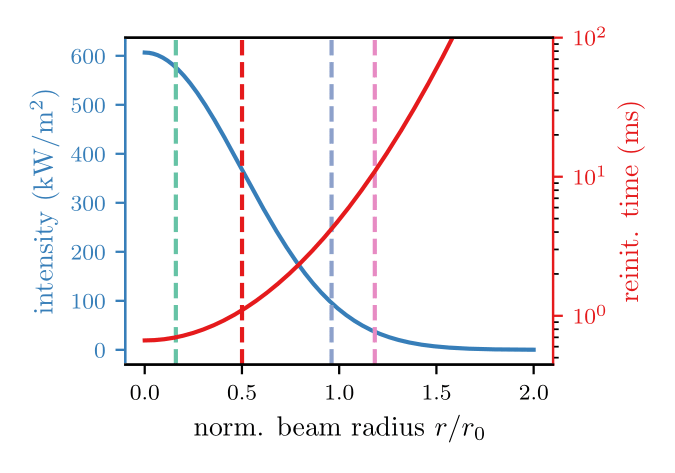}}
\caption[]{Modelled variation in laser intensity and the time to reinitialize NV centers into the $m_s$=0 ground state level as a function of bean radius, relative to $r_0$ the 1/e$^2$ beam width. The reinitialization time increases rapidly at the lower intensity edges of the beam.}
\label{fig:hyst-beamrad}
\end{figure}

\begin{figure}[htbp]
\center
\includegraphics[width=8.6cm]{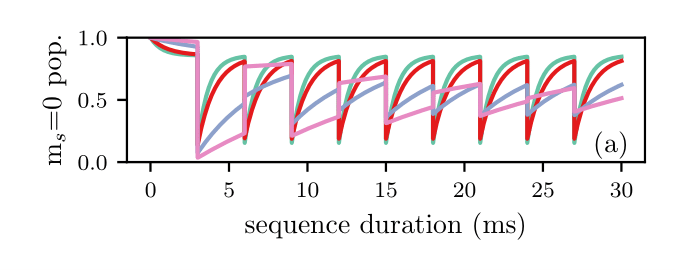}
\includegraphics[width=8.6cm]{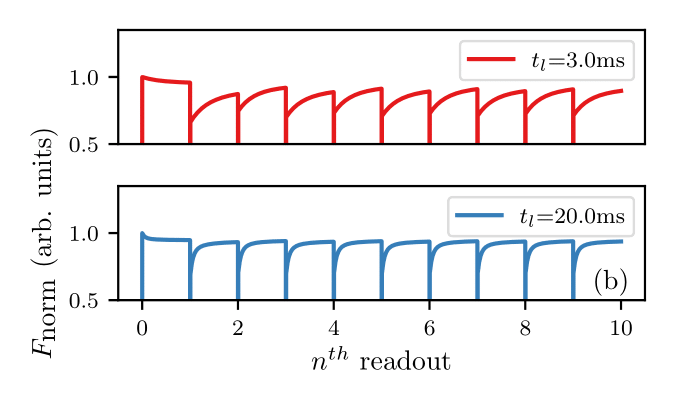}
\includegraphics[width=8.6cm]{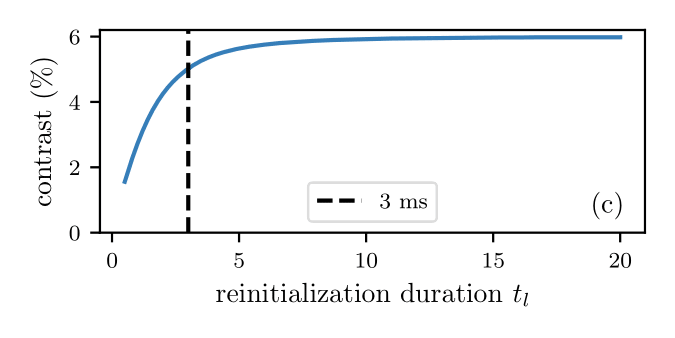}
\caption[]{(a) Modeled dynamics of the $m_s$=0 population for NV centers receiving pump beam intensity $I(r)$ at 4 different increasing values of $r$/$r_0$. (b) The hysteresis-free behaviour of the defects within $r$/$r_0<0.5$ dominates the fluorescence output (normalized to the initial output with all NV centers in the $m_s$=0 bright state). The NV centers at the beam edge that are not fully reinitialized reduce the ensemble contrast as laser pulse length is reduced, as modeled in (c).}
\label{fig:hyst-mainresults}
\end{figure}

We note that the fact we can achieve the same hysteresis-free contrast for a short \SI{3}{ms} laser pulse as for one much longer is somewhat surprising. We consider that this effect primarily arises due to the Gaussian intensity profile of the laser, whereby the NV centers at the low intensity edges of the beam require more time to reinitialize back into the ground state, but contribute far less to the overall fluorescence output, especially in the first few milliseconds of the readout laser pulse where contrast is measured\cite{Wolf2015}. In order to further investigate the physics of our NV ensemble and to determine the size of ensemble we address, we implemented a simple physical model of the NV population dynamics. Our model consists of a fixed NV density addressed by a radially (Gaussian-) varying laser beam intensity, with an NV at radius $r$ receiving a pump intensity $I(r)$. We then solve a rate model for all NV centers,\cite{Robledo2011, Dumeige2019} from which we estimate the relative fluorescence output and ensemble contrast $C$. We perform our microwave pulses as an ideal $\pi$-pulse with instantaneous population transfer in the rate model between levels $m_s$=0 to $m_s$=-1. Further details of the implementation of the model are given in Supplementary Information.

In Fig.~\ref{fig:hyst-beamrad}, we plot the relative intensity $I(r)$ and the reinitialization time, the exponential decay time required for the pump beam to return all NV centers at $r$ into the $m_s$=0 ground state as a function of beam radius $r$/$r_0$. Here $r_0$ represents the 1/e$^2$ beam width as in our experiment. From this simulation, it is clear that the time period over which we integrate to derive the experimental contrast ($t_w$=0.3-\SI{2.7}{ms}) corresponds to near complete reinitialization of the NV centers within $r$/$r_0\approx0.5$, or \SI{25}{\%} of the ensemble. Although this does not represent the entire ensemble, this still corresponds to  $\approx 1$ billion NV centers, based on estimated ensemble size (4 $\times$ 10$^9$) from our experimental measurement of fluorescence emission. 

The reinitilization dynamic behavior can be seen in Fig.~\ref{fig:hyst-mainresults}(a), plotting the time evolution of the $m_s$=0 state population for the first 10 readout/MW pulses of length $t_l=\SI{3}{ms}$ for 4 increasing values of $r$/$r_0$. Below $r$/$r_0=0.5$, hysteresis-free behaviour can be achieved in our model almost immediately after the first microwave pulse. Hysteretic behaviour is observed for NV centers further towards the edge of the beam. For these NV centers, the ground state occupancy decays to $\approx\SI{50}{\%}$ within the first 10-20 pulses. These NV centers therefore contribute by a reduced amount to the contrast (as measured in $t_w$) as compared to the NV centers in the beam centre, which exhibit the correct dynamics of full reinitialization by the laser and full state transfer by the microwave $\pi$-pulse. Since the outer NV centers are not fully reinitialized into the triplet ground state, they also act to produce a lower fluorescence emission as compared to the level expected with all NV centers reinitialised into the spin triplet ground state. This can be seen in Fig.~\ref{fig:hyst-mainresults}(b), showing the total ensemble fluorescence emission as a function of time for $t_l=\SI{20}{ms}$ and $t_l=\SI{3}{ms}$ laser pulses. For the shorter pulse length, a greater number of NV centers are not fully reinitialized, reducing the the overall fluorescence emission to approximately 90$\%$ of the maximum reached for $t_l$=20ms. Since the NV centers on the beam edge are not properly reinitialised into the ground state, they also cannot be correctly manipulated by the microwave state transfer pulse. As laser pulse length $t_l$ is reduced and this fraction of NV centers not fully reinitialised increases, the effect is therefore a reduction in contrast $C$ following a microwave pulse. This modelled behaviour can be seen in Fig.~\ref{fig:hyst-mainresults}(c) which qualitatively replicates our experimental data in Fig.~\ref{fig:contrastvsinit}. 

The ability to rapidly read and reinitialize in this manner is an extremely useful result, since it gives a means to adequately control and read a large NV ensemble with shorter laser pulses than that required to fully reinitialize every defect center. This significantly increases the measurement bandwidth for pulsed quantum sensing schemes, while still addressing a large number of defects required to maximize sensitivity. 

In order to model fluorescence behavior matching the experimental exponential decay using a Gaussian beam profile and associated volume, it was necessary to set modelled pump beam intensity a factor of 6 less than the intensity estimated experimentally. We attribute this to two factors not included in our model: reflection loss due to imperfect Brewster's angle coupling into the diamond and internal reflection within the diamond, spreading the beam across a wider volume of NV centers. The estimates of ensemble volume and number of defects addressed by the pump laser thus represents a lower bound based on the assumptions of our model. Further model development and investigations beyond the focus of this work are required to explore these aspects further, including observing changes in decay time for the fluorescence readout as a function of beam incidence angle and using a non-Gaussian laser profile.

\subsection{ODMR Using Shaped Optimal Control Pulses}
\label{section:5}
\begin{figure}[htbp]
\centerline{\includegraphics[width=8.6cm]{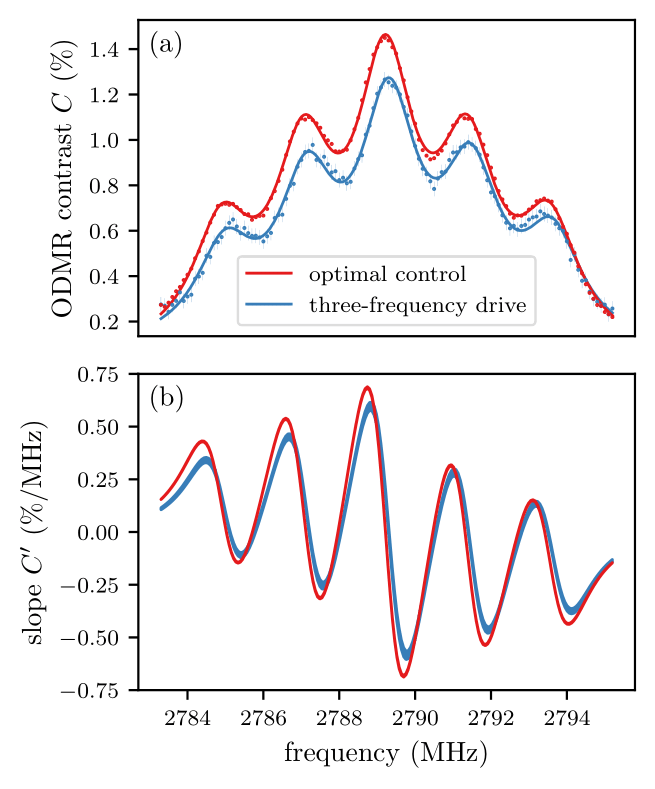}}
\caption[]{Comparison of pulsed ODMR measurements using the most sensitive optimized shaped pulse and the flat pulse that delivers the highest contrast using three-frequency drive. The slope data shown in (b) is the slope of the fit to the ODMR data in (a).}
\label{fig:fullodmrcompare}
\end{figure}

\begin{figure}[htbp]
\centerline{\includegraphics[width=8.6cm]{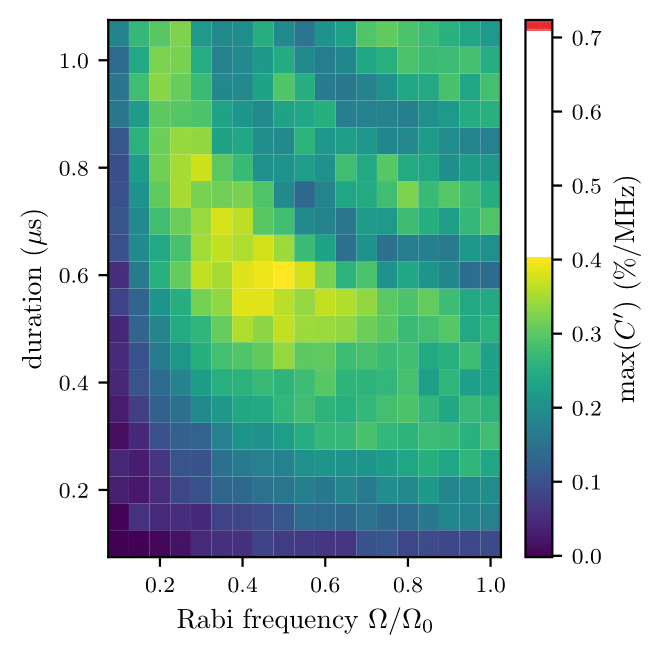}}
\caption[]{The maximum contrast slope $C^{'}$ measured for flat single-frequency drive pulses over the relevant parameter space of microwave power (plotted as Rabi frequency) and duration. The red stripe in the colorbar shows the maximum contrast slope of the best optimal control pulse from Fig.~\ref{fig:fullodmrcompare}. The equivalent plot for 3 frequency drive is given in the Supplementary Information.}
\label{fig:2d1fd}
\end{figure}

\begin{figure}[htbp]
\centerline{\includegraphics[width=8.6cm]{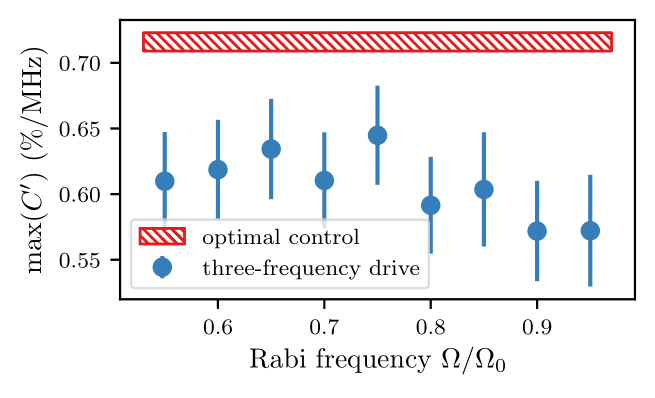}}
\caption[]{ODMR spectrum frequency versus contrast slope C$^{'}$ for the shaped optimal control pulse and for the best flat 3-frequency drive pulses of any duration for a given microwave power (plotted as measured Rabi frequency). Error bars and the y-range of the optimal control pulse represent 1$\sigma$ uncertainty. The power used for the optimal control pulse was 0.91 of the maximum Rabi frequency $\Omega_0$. The plot covering the full parameter space in time and Rabi frequency is given in Supplementary Information.}
\label{fig:flatcompare}
\end{figure}

Using our optimal control algorithm including all 3 hyperfine levels for \ce{^{14}N}, we first calculated a series of shaped microwave pulses spanning the parameter space of $\hat{\Delta}$ and $\hat{\alpha}$, the Rabi frequency limit $R_\text{lim}$, and the pulse duration $t_p$. Their performance was then tested experimentally to explore the limits of these parameters that yield high contrast $C$ and ODMR slope $C^{'}$.  We found that extending $\hat{\Delta}$ and $\hat{\alpha}$ beyond $\pm \SI{1}{MHz}$ and $\pm \SI{10}{\%}$ respectively had negligible impact, likely indicating that the real ensemble distribution in our diamond was within these ranges. Having found that pulses in the range of $\SI{1.1}{MHz} < R_\text{lim} < \SI{2.4}{MHz}$ and $\SI{1}{\micro\second} < t_p < \SI{2}{\micro\second}$ performed well, we experimentally searched the parameter space of these optimal control pulses applied by producing ODMR spectra using the shaped pulses and searching for the maximum slope $C^{'}$. We found the best-performing pulse optimized with $t_{p}=\SI{1.85}{\micro\second}$ and $R_\text{lim}=\SI{1.4}{MHz}$, with similar performance from larger $R_\text{lim}$ up to 2~MHz at the same $R_\text{exp}$. The modulation components $I(t)$ and $Q(t)$ for this pulse are shown in Fig.~\ref{fig:IQ}(a), and the control amplitudes are given in the Supplementary Information. 

The experimental ODMR spectrum from the best shaped control pulse found is shown in Fig.~\ref{fig:fullodmrcompare}(a). By differentiating the spectrum, we also show the frequency versus contrast slope $C^{'}$ in Fig.~\ref{fig:fullodmrcompare}(b). Here the largest possible slope is desired, since this produces the maximum response and highest sensitivity. For comparison, we plot in the same figure the ODMR spectrum using a flat three-frequency drive ($\pi$-)pulse that corresponds to the maximum slope for a conventional pulse without shaping. We found the maximum slope to be 11$\%$ higher for the shaped optimal control pulse than for this conventional flat pulse. Compared to the simplest single-frequency flat ($\pi$-)pulses most used in literature, we found significant improvement of up to 73\%. This corresponds directly to the same factor of improvement in sensitivity. 

We note that the length of the flat and shaped pulse that delivered maximum slope were significantly different. This could potentially lead to the longer shaped pulse achieving higher performance purely by delivering more microwave power over an extended time period. To ensure this was not the case, we compared the optimised pulse against single and three-frequency drive flat pulses over an extended parameter space of pulse lengths (up to $t_p=\SI{1.35}{\micro\second}$) and applied microwave power (up to Rabi frequency $R_{max}=\SI{3}{MHz}$) . This data is shown in Fig.~\ref{fig:2d1fd} for single frequency drive and Fig.~\ref{fig:flatcompare} for three-frequency drive. The flat pulses performed best at the length and power that corresponded to performing a $\pi$-pulse on the largest possible subset of NV centers (maximizing contrast). However, as can be seen from these figures, the shaped microwave pulse we created using our optimal control methods always produced an ODMR slope far higher than any unshaped drive. This was the case for any pulse length or microwave power, with the optimum for the flat pulses reached well within experimental limits of $R_\text{max}$ and $t_p$.

For our setup, we can estimate shot-noise-limited sensitivity using the expression derived in Appendix~\ref{app:sensitivity}: 

\begin{equation*}
 \eta \approx \frac{\sqrt{2t_{R}t_{I}}}{\gamma_{e}C'\tau_{R}(1 - e^{-t_{R}/\tau_{R}})\sqrt{R_{0}}},
\end{equation*}
where we take into account the times for readout and reinitialization $t_{R}, t_{I}$, the reinitialization decay constant $\tau_{R}$, photon collection rate at max power $R_{0}$, electron gyromagnetic ratio $\gamma_{e}$, and measured contrast slope $C'$. For our setup, we estimate $\eta \approx \SI{10}{\nano\tesla/\sqrt{\hertz}}$. Although this is lower than state of the art figures reported elsewhere for magnetic field sensing with NV centers, we note that our setup is not optimized for sensitivity due to the standard optical grade diamond we use (rather than one materially optimized for sensing), our small APD detector area and ADC memory limitations. 

\section{Conclusion}

In this work, we demonstrate that a large ensemble of solid state defects in a macroscopic sample can be manipulated and coherently controlled in a manner beneficial for quantum sensing. We demonstrate this for an ensemble of NV centers in diamond through the use of  shaped microwave pulses generated  using Floquet theory and optimal control methods. Due to the scaling of sensitivity with the number of defects, such large ensembles are key for quantum sensing applications, either using NV centers or other solid state defects. Both our overall NV ensemble volume within the estimated Gaussian beam width ($\approx$ 4 $\times$ 10$^{9}$ NV centers in a  $\approx\SI{0.04}{mm^3}$ volume) and our estimated NV ensemble contributing maximally to the contrast signal ($\approx \SI{25}{\%}$ of the total) was larger than NV ensembles previously studied and reported in the literature using optimal control methods largely studied using confocal microscopy.\cite{Bjorn2015, Nobauer2015a, Nobauer2015, Bartels2013, Hernandez-Gomez2019, Konzelmann2018, Dong2019}

By fully considering the physics of the defect system and including the hyperfine interaction in our optimization, we demonstrate an \SI{11}{\%} enhancement in ODMR slope with optimized shaped pulses when compared to the best alternative 3-frequency drive flat (fixed amplitude and phase) ($\pi$-)pulses and a \SI{78}{\%} improvement over standard single-frequency-drive flat ($\pi$-)pulses most commonly used for coherent control in the literature. These are directly equivalent to the same factor of sensitivity improvement when used in an applied sensing scheme. This significant improvement offers potential for wider impact for DC/low-frequency sensing, for example in precision measurement of slowly varying temperature where ensemble probe bandwidth limitations imposed by the $\approx\SI{5}{\micro\second}$ shaped pulse length would be less constraining. 

We estimate a shot noise-limited sensitivity of \SI{10}{nT/\sqrt{Hz}} using our setup, while noting that neither the diamond we use nor our apparatus was optimized to maximize sensitivity at this time. Our method is not specific to the apparatus we used and could be applied equally well to a sensitivity-optimized setup, for example using an isotopically purified diamond. By measuring the ODMR contrast by referring to the signal from an additional photodetector, we were able to reject more of the laser technical noise while maximizing the number of contrast measurements we could achieve as compared to alternative time domain noise rejection methods \cite{Wolf2015}. 

Through modeling of the physical dynamics of the readout and initialization of the defect ensemble, we show that although many tens of milliseconds are required to fully reinitialise the whole NV ensemble, a shorter laser pulse can address and reinitialise a large proportion of the NV centers. By demonstrating reliable contrast measurements free of hysteresis, we show that these NV centers can be addressed and controlled reliably. Further work is required to fully understand the dynamics of the system and the distribution of pump light in the diamond. However, our measurements suggest the primarily limiting factor on the readout is the Gaussian shape of the laser beam, hinting at considerable future improvement using a non-Gaussian profile.

The shaped microwave pulses we generate in this work almost certainly represent local maxima of performance in a wide parameter space. We consider it very likely that advances in methods for optimization as well as experimental improvements could provide even better solutions in future. A particular flaw is the assumption of simple Gaussian distributions for detuning and other parameters, which are a poor representation of the actual properties of a real sample.  A route forward may be to use experimental feedback in the optimization algorithm. This would be simplified by producing a more homogeneous microwave field through antenna improvements, increasing the ensemble Rabi frequency through better use of the microwave power, and the use of alternative laser beam profiles to improve uniformity of initialization and readout. Additionally, in this work we optimize for state transfer $\ket{0}$ to $\ket{-1}$, which aims to maximize contrast $C$. By instead explicitly optimizing for the change in contrast in response to the control field (the slope $C^{'}$ in our results above), better optimized pulses could be generated.

Our work represents an important step in the direction of using optimal control and other techniques widely used in nuclear magnetic- and electron spin- resonance experiments to explore the physics of new systems suitable for quantum sensing. These techniques, including those we outline here, can be adapted to be widely applicable, not only to diamond but to other defects in both bulk and novel quantum materials, such as those in 2D materials.\cite{Gottscholl2020} Using control pulses shaped by optimal control methods, which could be either microwaves, optical fields or some other means, offers the best route to reach the ultimate $T_{2}^{*}$-limited sensitivity for any suitable quantum system. 

\section{Acknowledgements}
The work presented here was funded by the Novo Nordisk foundation through the synergy grant bioQ and the bigQ Center funded by the Danish National Research Foundation (DNRF).

\begin{appendices}
\section{Spin Matrices} \label{App:spin}
Below are shown the 6-by-6 matrix representations of the Pauli spin matrices that are each specific to one of the three nitrogen-14 hyperfine transitions.
\begin{equation}
    \sigma_{z, 1} = \left( \begin{smallmatrix} 1 & 0 & 0 & 0 & 0 & 0 \\ 0 & -1 & 0 & 0 & 0 & 0 \\ 0 & 0 & 0 & 0 & 0 & 0 \\ 0 & 0 & 0 & 0 & 0 & 0 \\ 0 & 0 & 0 & 0 & 0 & 0 \\ 0 & 0 & 0 & 0 & 0 & 0 \end{smallmatrix} \right), \:
    \sigma_{z, 2} = \left( \begin{smallmatrix} 0 & 0 & 0 & 0 & 0 & 0 \\ 0 & 0 & 0 & 0 & 0 & 0 \\ 0 & 0 & 1 & 0 & 0 & 0 \\ 0 & 0 & 0 & -1 & 0 & 0 \\ 0 & 0 & 0 & 0 & 0 & 0 \\ 0 & 0 & 0 & 0 & 0 & 0 \end{smallmatrix} \right)
\end{equation}
\begin{equation}
    \sigma_{z, 3} = \left( \begin{smallmatrix} 0 & 0 & 0 & 0 & 0 & 0 \\ 0 & 0 & 0 & 0 & 0 & 0 \\ 0 & 0 & 0 & 0 & 0 & 0 \\ 0 & 0 & 0 & 0 & 0 & 0 \\ 0 & 0 & 0 & 0 & 1 & 0 \\ 0 & 0 & 0 & 0 & 0 & -1 \end{smallmatrix} \right), \:
    \sigma_{x, 1} = \left( \begin{smallmatrix} 0 & 1 & 0 & 0 & 0 & 0 \\ 1 & 0 & 0 & 0 & 0 & 0 \\ 0 & 0 & 0 & 0 & 0 & 0 \\ 0 & 0 & 0 & 0 & 0 & 0 \\ 0 & 0 & 0 & 0 & 0 & 0 \\ 0 & 0 & 0 & 0 & 0 & 0 \end{smallmatrix} \right)
\end{equation}
\begin{equation}
    \sigma_{x, 2} = \left( \begin{smallmatrix} 0 & 0 & 0 & 0 & 0 & 0 \\ 0 & 0 & 0 & 0 & 0 & 0 \\ 0 & 0 & 0 & 1 & 0 & 0 \\ 0 & 0 & 1 & 0 & 0 & 0 \\ 0 & 0 & 0 & 0 & 0 & 0 \\ 0 & 0 & 0 & 0 & 0 & 0 \end{smallmatrix} \right), \:
    \sigma_{x, 3} = \left( \begin{smallmatrix} 0 & 0 & 0 & 0 & 0 & 0 \\ 0 & 0 & 0 & 0 & 0 & 0 \\ 0 & 0 & 0 & 0 & 0 & 0 \\ 0 & 0 & 0 & 0 & 0 & 0 \\ 0 & 0 & 0 & 0 & 0 & 1 \\ 0 & 0 & 0 & 0 & 1 & 0 \end{smallmatrix} \right)
\end{equation}
\begin{equation}
    \sigma_{y, 1} = \left( \begin{smallmatrix} 0 & -i & 0 & 0 & 0 & 0 \\ i & 0 & 0 & 0 & 0 & 0 \\ 0 & 0 & 0 & 0 & 0 & 0 \\ 0 & 0 & 0 & 0 & 0 & 0 \\ 0 & 0 & 0 & 0 & 0 & 0 \\ 0 & 0 & 0 & 0 & 0 & 0 \end{smallmatrix} \right), \:
    \sigma_{y, 2} = \left( \begin{smallmatrix} 0 & 0 & 0 & 0 & 0 & 0 \\ 0 & 0 & 0 & 0 & 0 & 0 \\ 0 & 0 & 0 & -i & 0 & 0 \\ 0 & 0 & i & 0 & 0 & 0 \\ 0 & 0 & 0 & 0 & 0 & 0 \\ 0 & 0 & 0 & 0 & 0 & 0 \end{smallmatrix} \right)
\end{equation}
\begin{equation}
    \sigma_{y, 3} = \left( \begin{smallmatrix} 0 & 0 & 0 & 0 & 0 & 0 \\ 0 & 0 & 0 & 0 & 0 & 0 \\ 0 & 0 & 0 & 0 & 0 & 0 \\ 0 & 0 & 0 & 0 & 0 & 0 \\ 0 & 0 & 0 & 0 & 0 & -i \\ 0 & 0 & 0 & 0 & i & 0 \end{smallmatrix} \right)
\end{equation}

\section{Estimation of shot noise-limited sensitivity}
\label{app:sensitivity}

Shot noise-limited sensitivity estimation is typically\cite{Dreau2011,OptimisedFmod} similar to
\begin{equation*}
\eta \approx \frac{1}{\gamma_{e}\sqrt{R}C'}
\end{equation*}
with the electron gyromagnetic ratio $\gamma_{e}$, the photon detection rate $R$ and the (in this case empirically measured) ODMR slope $C'=\dv{C}{f}$.
This assumes that each collected photon adds the same amonut of information, which is the case in a typical pulsed detection setup where the readout time $t_{R}$ is much shorter than the total reinitialization time $t_{I}$, and the contrast barely decays during $t_{R}$.
We therefore include a factor representing the mean information collected per photon
\begin{equation*}
  \frac{1}{t_{R}}\int_{0}^{t_{R}}e^{-t/\tau_{R}}\dd t = \frac{\tau_{R}(1 - e^{-t_{R}/\tau_{R}})}{t_{R}}
\end{equation*}
with $\tau_{R}\approx\SI{1.4}{ms}$ as the decay constant of the contrast, making this factor about 0.5. We additionally modify $R$ for clarity, in terms of the maximum photon collection rate at a peak of the modulation $R_{0}$, $R=R_{0}t_{R}/2t_{I}$, where the duration of the RF pulse is neglected, and the $1/2$ results from the modulation. In all, we obtain
\begin{equation*}
 \eta \approx \frac{\sqrt{2t_{R}t_{I}}}{\gamma_{e}C'\tau_{R}(1 - e^{-t_{R}/\tau_{R}})\sqrt{R_{0}}}.
\end{equation*}

\end{appendices}

\bibliographystyle{unsrt}
\bibliography{pulsebib2}

\end{document}